\documentclass{elsart}
\usepackage{epsfig}       
\begin{document}
\runauthor{Kri\v zan}
\begin{frontmatter}
\title{The HERA-B Ring Imaging \v Cerenkov Counter
}

\vspace{-1cm}

\center{\author{}}\\
\thanks{\hspace{1mm}
Now at Max-Planck-Institut f\" ur Plasmaphysik Garching/Greifswald, Germany}
\thanks{\hspace{1mm}
Now at University of Rostock, Rostock, Germany}
\thanks{\hspace{1mm}
Permanent address: Slovenian Scientific Institute, Klagenfurt, Austria}
\thanks{\hspace{1mm}
Now at General Atomics, San Diego, USA}
\thanks{\hspace{1mm}
Now at Lawrence Berkeley National Lab, Berkeley, CA, USA}
\author[UB]{I.~Ari\~no,}
\author[CO]{J.~Bastos,}
\author[NU]{D.~Broemmelsiek,}
\author[CO]{J.~Carvalho,}
\author[UB]{M.~Chmeissani,}
\author[UB]{P.~Conde,}
\author[UH]{J.~Davila,}   
\author[UTA]{D.~Dujmi\' c,}
\author[UTA]{R.~Eckmann,}
\author[UB]{L.~Garrido,}
\author[UB]{D.~Gascon,}
\author[UTA]{T.~Hamacher,}$^{,1}$
\author[IJS]{A.~Gori\v sek,}
\author[CO]{I.~Ivaniouchenkov,}
\author[UH]{M.~Ispirian,}
\author[DESY]{S.~Karabekian,}$^{,2}$
\author[UTA]{M.~Kim,}
\author[MB,IJS]{S.~Korpar,}
\author[FMF,IJS]{P.~Kri\v zan,}
\author[IJS]{S.~Kupper,}$^{,3}$
\author[UH]{K.~Lau,}
\author[NU]{P.~Maas,}
\author[UTA]{J.~McGill,}$^{,4}$
\author[UB]{R.~Miquel,}$^{,5}$
\author[UH]{N.~Murthy,}   
\author[UB]{D.~Peralta,}
\author[IJS]{R.~Pestotnik,}
\author[UH]{J.~Pyrlik,}
\author[UH]{S.~Ramachandran,}
\author[UTA]{K.~Reeves,}
\author[NU]{J.~Rosen,}
\author[UHH]{W.~Schmidt-Parzefall,}
\author[DESY]{A.~Schwarz,}
\author[UTA]{R.F.~Schwitters,}
\author[UB]{X.~Siero,}
\author[IJS]{M.~Stari\v c,}
\author[FE,IJS]{A.~Stanovnik,}
\author[IJS]{D.~\v Skrk,}
\author[IJS]{T.~\v Zivko}
\address[UTA]{University of Texas, Austin, USA}
\address[UB]{University of Barcelona, Spain}
\address[CO]{LIP Coimbra, Portugal}
\address[NU]{Northwestern University, Evanston, USA}
\address[DESY]{DESY, Hamburg, Germany}
\address[UHH]{University of Hamburg, Hamburg, Germany}
\address[UH]{University of Houston, Houston, USA}
\address[FE]{Faculty of Electrical Engineering, University of Ljubljana,
Slovenia}
\address[IJS]{J.~Stefan Institute, Ljubljana, Slovenia}
\address[FMF]{Faculty of
Mathematics and Physics, University of Ljubljana,  Slovenia}
\address[MB]{Faculty of Chemistry and Chemical
Engineering, University of Maribor, Slovenia}

\begin{abstract}

The HERA-B RICH uses a radiation path length of
2.8~m in
C$_4$F$_{10}$ 
gas and a large 24~m$^2$ spherical mirror for
imaging \v Cerenkov rings. The photon
detector consists of 2240 Hamamatsu multi-anode photomultipliers with 
about 27,000 channels.
A 2:1 reducing two-lens telescope in front of each PMT
increases the sensitive area at the expense of increased pixel size,
resulting in a
contribution to the resolution which roughly matches that of dispersion.
The counter was completed in January of 1999, and its  performance  
 has been steady and reliable over the years it has been in operation.
The design performance of the RICH was fully reached: the average 
number of detected photons in the RICH for a $\beta = 1$ particle was
found to be 33 with a single hit resolution of 0.7 mrad and 1 mrad in
the fine and coarse granularity regions, respectively.

\end{abstract}
\end{frontmatter}

\section{Introduction}
\begin{figure}[hbt]
\centerline{\epsfig{file=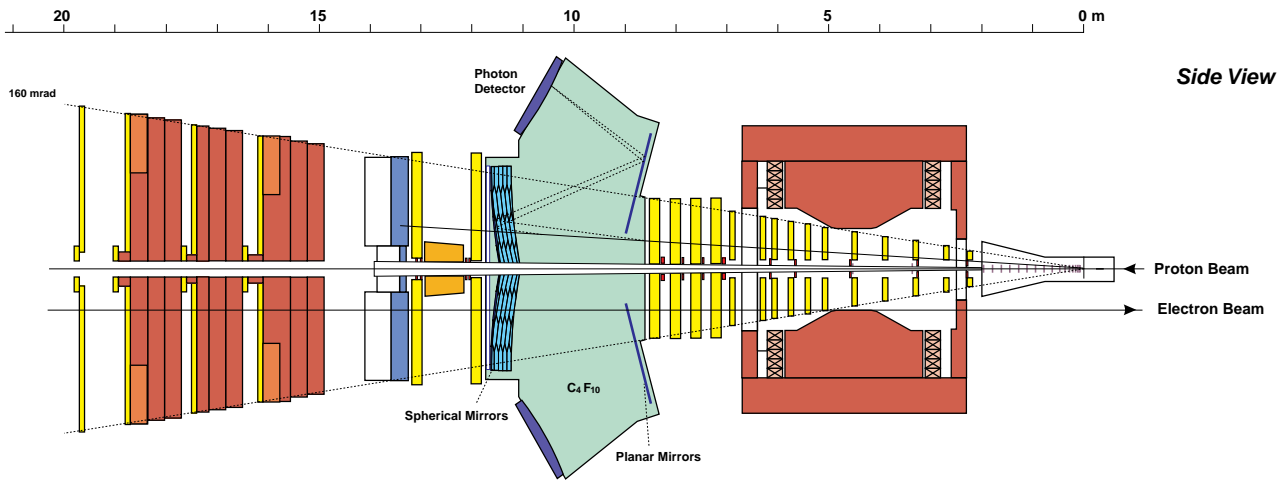,height=17.cm,angle=-90,clip}}
\caption[kk]{A side  view of the HERA-B detector.
}
\label{hbdet}
\end{figure}

HERA-B, a fixed target experiment
(see Fig.~\ref{hbdet})
at the HERA storage ring  at DESY,
was designed \cite{prop}  to  measure
rare processes in the decays of $B$ mesons. The  $B$ mesons
are produced in collisions
 of  920 GeV/c protons  with a  fixed target, which consists of 
8 wires which can be individually inserted 
into the halo of the proton beam in order not to disturb experiments 
measuring $ep$ collisions.
One of the essential components of the spectrometer is the 
Ring Imaging Cherenkov counter (RICH)
\cite{prop,msm,rich-beauty97,korpar-rich98}. 
The main purpose of the RICH  counter 
is the identification of charged hadrons, in particular kaons from decays
 of   $B$ mesons. 
Identifying charged kaons essentially means separating them 
from pions in the momentum range
between 3 GeV/c and about 50 GeV/c at an interaction rate of 
up to 40 MHz.

This paper is organized as follows. We shall first discuss the
design criteria and their implementation. The components of the counter
will be presented, as well as the monitoring and analysis programs.
Finally,  results of measurements made with the counter will be
discussed, and the performance of the RICH as an identification
device, and also as a tracking device, will be presented.

\section{Design Criteria}

In a fixed target experiment such as HERA-B where
one has to deal with an intense flux of charged particles, 
the design choices
of a ring imaging \v Cerenkov counter 
are governed by the following criteria.
To achieve the necessary performance, a sufficient number of \v Cerenkov
photons (at least 20) has to be detected per  ring image
of a $\beta=1$ particle~\cite{prop}. 
This requirement fixes the length of a gas radiator to a few meters.
The high rate capability of the photon detector requires a
detector with pad readout, which
limits the resolution to a few  millimeters.
This, in turn, together with the required high resolution in the 
measurement of the \v Cerenkov angle for single photons
(better than 1~mrad)
sets the  mirror focal length value to  several meters.
Last but not least,
the requirement that single photons are efficiently detected
with low background, forces
the photon detector 
to be kept out of the solid angle for charged particles and
conversion products.

\section{Implementation}

\subsection{Radiator}
%
Perfluorobutane gas (${\rm C_{4}F_{10}}$)
was 
chosen for the HERA-B RICH radiator,
since it combines a relatively high refractive index, 
$n = 1.00137$, and low dispersion
(5\% variation in $n-1$ 
over the  spectral acceptance from 300 nm to 600 nm~\cite{ullaland}).
In this gas the \v Cerenkov 
radiation threshold momenta for pions and kaons are 2.7 GeV/c and
9.6 GeV/c, respectively.
For ${\rm \beta =1}$ particles, the \v Cerenkov angle is 52.4~mrad, 
while the ${\rm \pi - K}$
difference in \v Cerenkov angle is 0.9~mrad at 50~GeV/c. The
r.m.s.~angular
spread due to dispersion amounts to 0.33~mrad.
Since some  freons are known to scintillate considerably, 
(e.g. CF$_4$), we have
studied  C$_4$F$_{10}$ scintillation~\cite{rok-diploma}
and found that this
contribution to background is less than 0.2 detected photons
per charged particle.

\begin{figure}[thb]
\centerline{\hbox{\psfig{figure=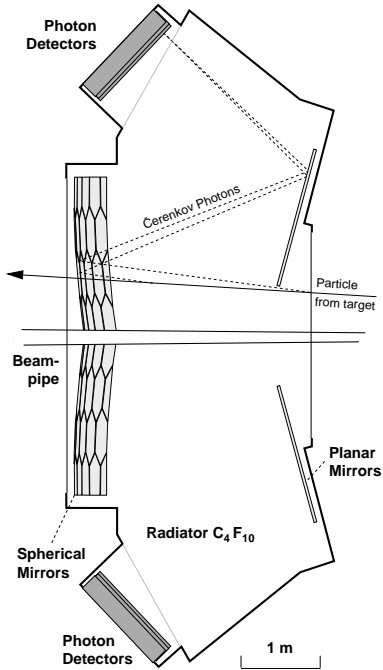,width=5cm}}}
\caption[kk]{Scheme of the RICH. Rays emitted by a particle and their 
	 paths to the photon detector are indicated.}
\label{rich-scheme}
\end{figure}

The radiator vessel, made from stainless steel plates with 1~mm aluminum
particle entrance and exit windows (see Fig.~\ref{rich-scheme}),
is placed about 8.5~m downstream of
the target (see Fig.~\ref{hbdet}).
Two beam shrouds made of carbon fiber-reinforced epoxy 
resin CFK close the gas volume around the two beam pipes for protons and electrons.
The \v Cerenkov
light exits the vessel through  2~mm
thick UV grade Plexiglass windows \cite{plexyglass}.
The vessel is filled with 108 m$^3$
of C$_4$F$_{10}$ gas, which
 is being circulated in a closed system with liquefaction
stages for cleaning and buffering \cite{gas-system,bosteels}.
Requirements on impurities are
moderate due to  operation in the visible and near UV 
part of the \v Cerenkov spectrum.
The radiator gas is kept at 2.5~mbar overpressure
relative to ambient pressure.
Additionally, a bubbler with a 10~mbar pressure difference threshold
is attached to the vessel.

\subsection{Mirrors}
\begin{figure}[thb]
\centerline{\hbox{\psfig{figure=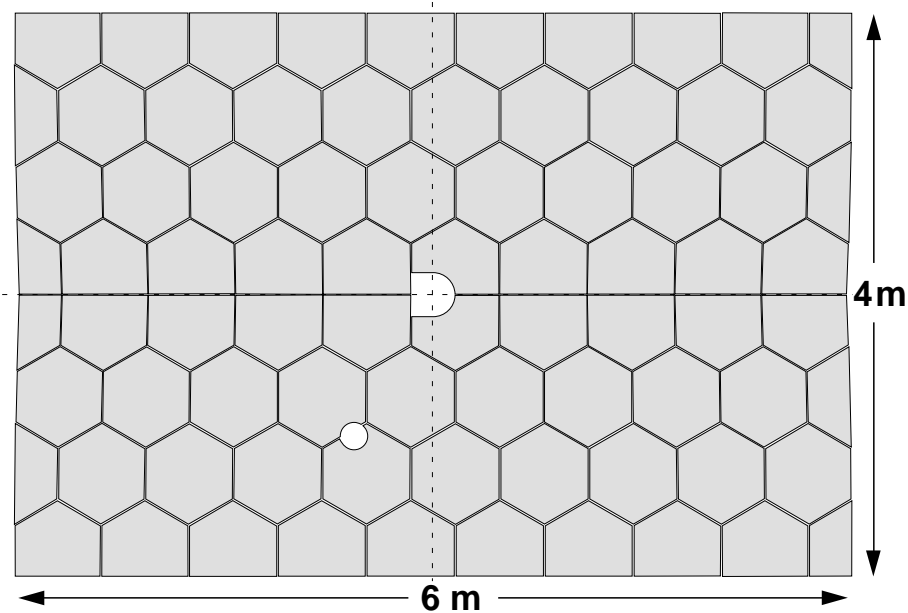,width=12cm}}}
\caption[kk]{Distribution of spherical mirror polygons. The holes in the
          array are for the proton and electron beam pipes.}
\label{mirror-tiling}
\end{figure}

The main imaging device is a spherical mirror placed inside the
radiator vessel with the center of the sphere near the target and a radius of
curvature of 11.4~m. The mirror, a 6~m by 4~m rectangular cutout from
the sphere, consists of 80 full or partial hexagons made
from 7~mm thick
Pyrex glass, coated with 200~nm of aluminum and 30 nm of MgF$_2$ (see
Fig. \ref{mirror-tiling}). To be able to place the focal surface 
outside the particle
flux ($\pm$160 mrad vertically), the mirror is split horizontally and
both halves are tilted by 9$^{\circ}$ away from the beam-line. A set
of two planar mirrors, composed of 18 rectangular elements each,
translates the focal surface to the photon detector area 
above and below the
radiator vessel (see Fig. \ref{rich-scheme}).
The planar mirrors are made of float glass, thus being significantly cheaper
than Pyrex mirrors at the required optical quality. 
All 80 spherical and 36
planar mirrors are mounted on a rigid, low mass support structure
inside the radiator volume and can be individually adjusted by stepper
motors from the outside. 

 The mirror quality was determined upon delivery 
by measuring  for each segment the
radius of curvature and the fraction of reflected light,
and by recording a Ronchi image to check the
homogeneity of the mirror surface~\cite{denis-phd}.
The reflectivity was required to exceed
85\% in the wavelength  interval from 250~nm to 600~nm. 
From  the data gathered, the  mirror segments were
grouped  in a tiling scheme according to their optical quality
and resolution requirements \cite{mirror-tilling}.

All mirrors were aligned to better than
the required precision by surveying them inside the vessel.
An off-line data based algorithm was developed for monitoring of
the positions of the mirror segments during the measurement
\cite{mirror-align}.

\subsection{Photon detector}

Two types of
wire chamber based photon detectors were initially considered,
 a CsI photocathode in a MWPC, and  a TMAE  detector with 10 cm deep,
8 x 8 mm$^2$ unit cells. After considerable success of on-the-bench
and beam tests \cite{tmae-rich95,csi-94,csi-ieee94}, the detectors were
tested in a high rate environment, as expected in the HERA-B experiment.
Both detectors had to be 
abandoned; the TMAE detector showed a prohibitive decrease of avalanche gain
due to aging effects \cite{tmae-ageing,tmae-heat}, 
while the CsI photocathode could not be routinely 
produced and maintained with sufficiently high quantum efficiency, in addition 
to problems with rates in excess of a few kHz per pixel 
\cite{csi-rich95,beaune96}. 

\begin{figure}[htb]
\centerline{\psfig{file=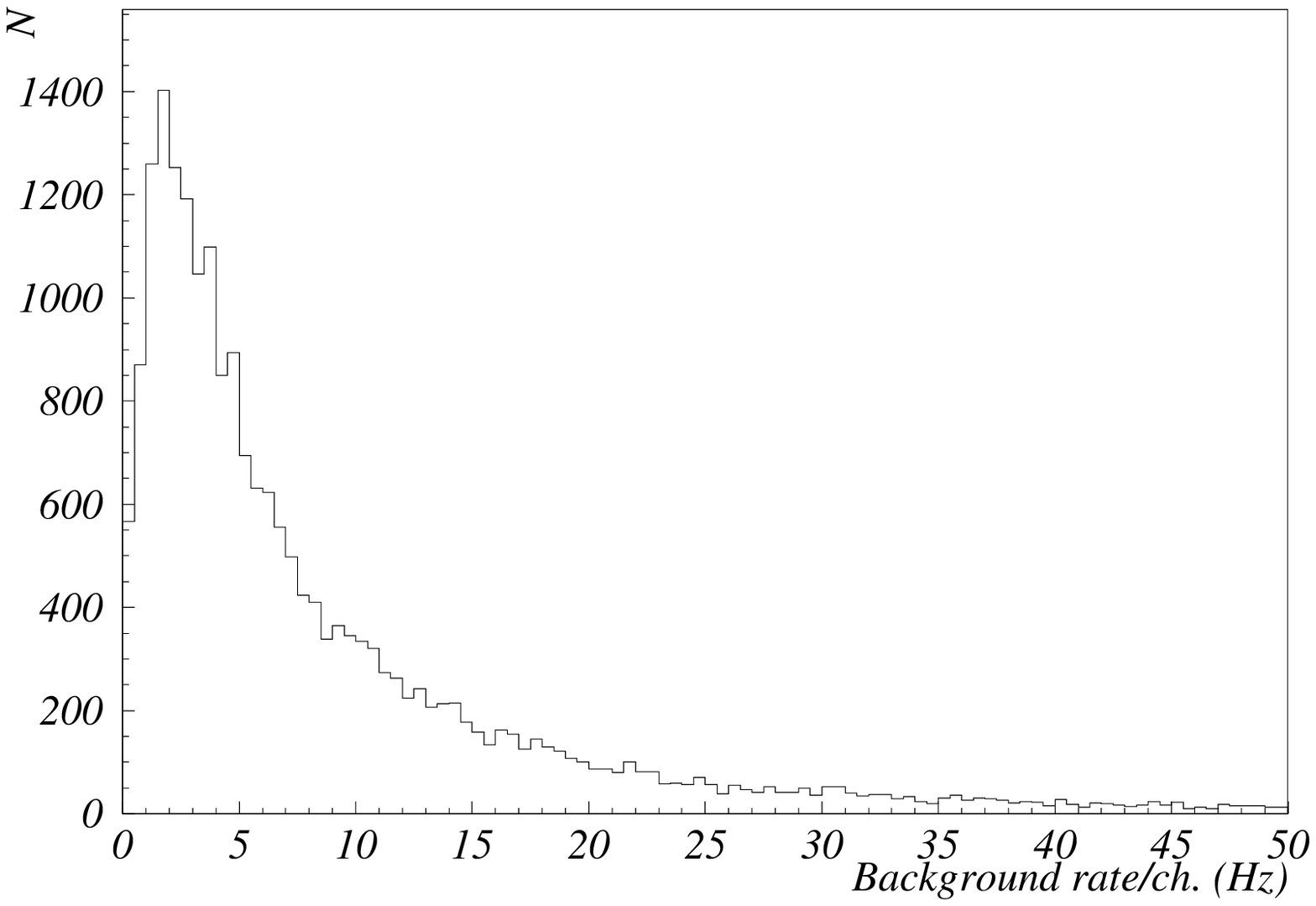,width=12cm}}

\vspace{-1cm}

\centerline{\psfig{file=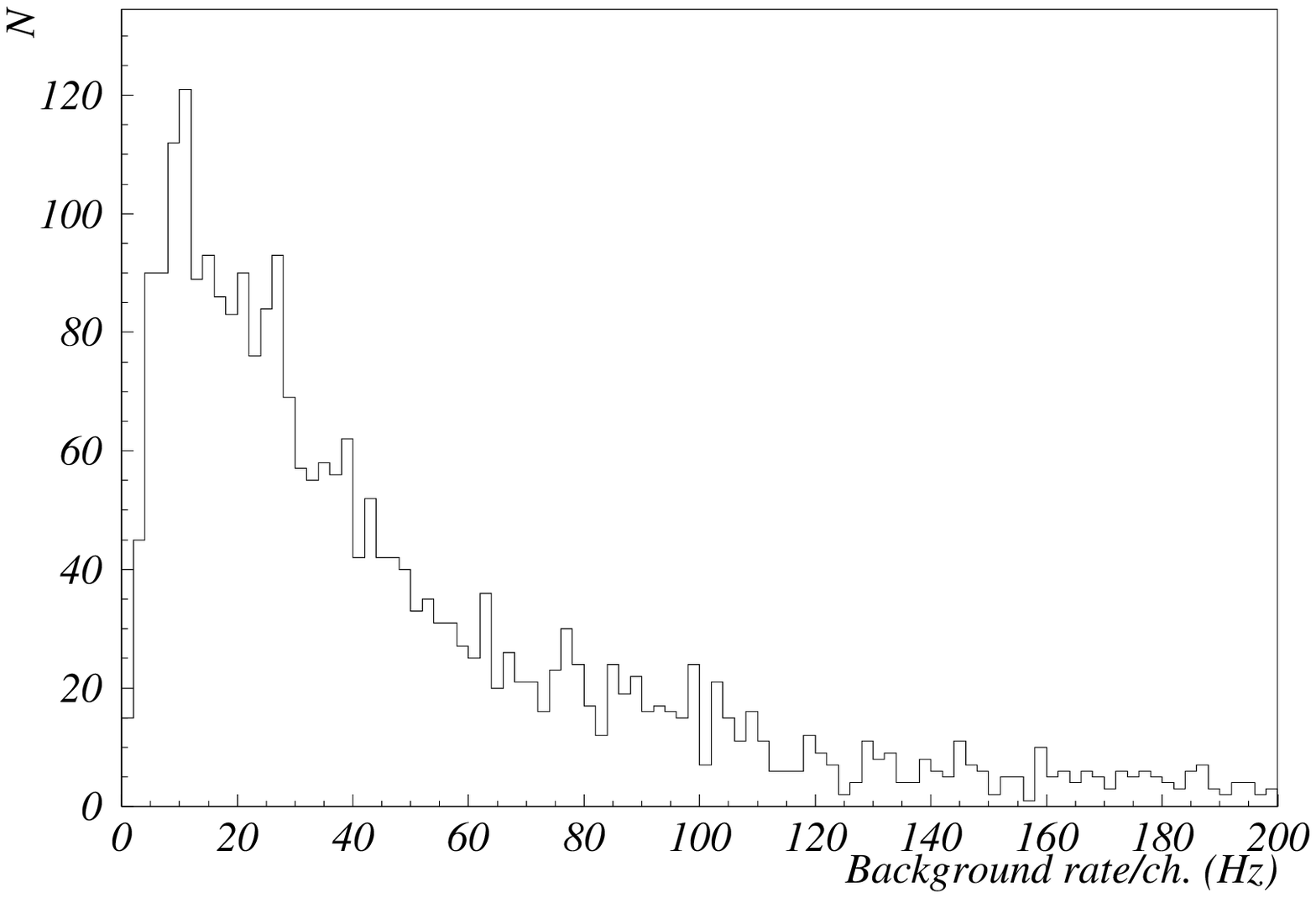,width=12cm}}
\caption[kk]{ Distribution of the number of PMT channels depending on the
background rate per channel for M16 (top) and M4 (bottom) PMTs, as measured after the tubes were 
kept in a light tight box for 15 minutes.
}
\label{figbg}
\end{figure}

\begin{figure}[htb]
\centerline{\psfig{file=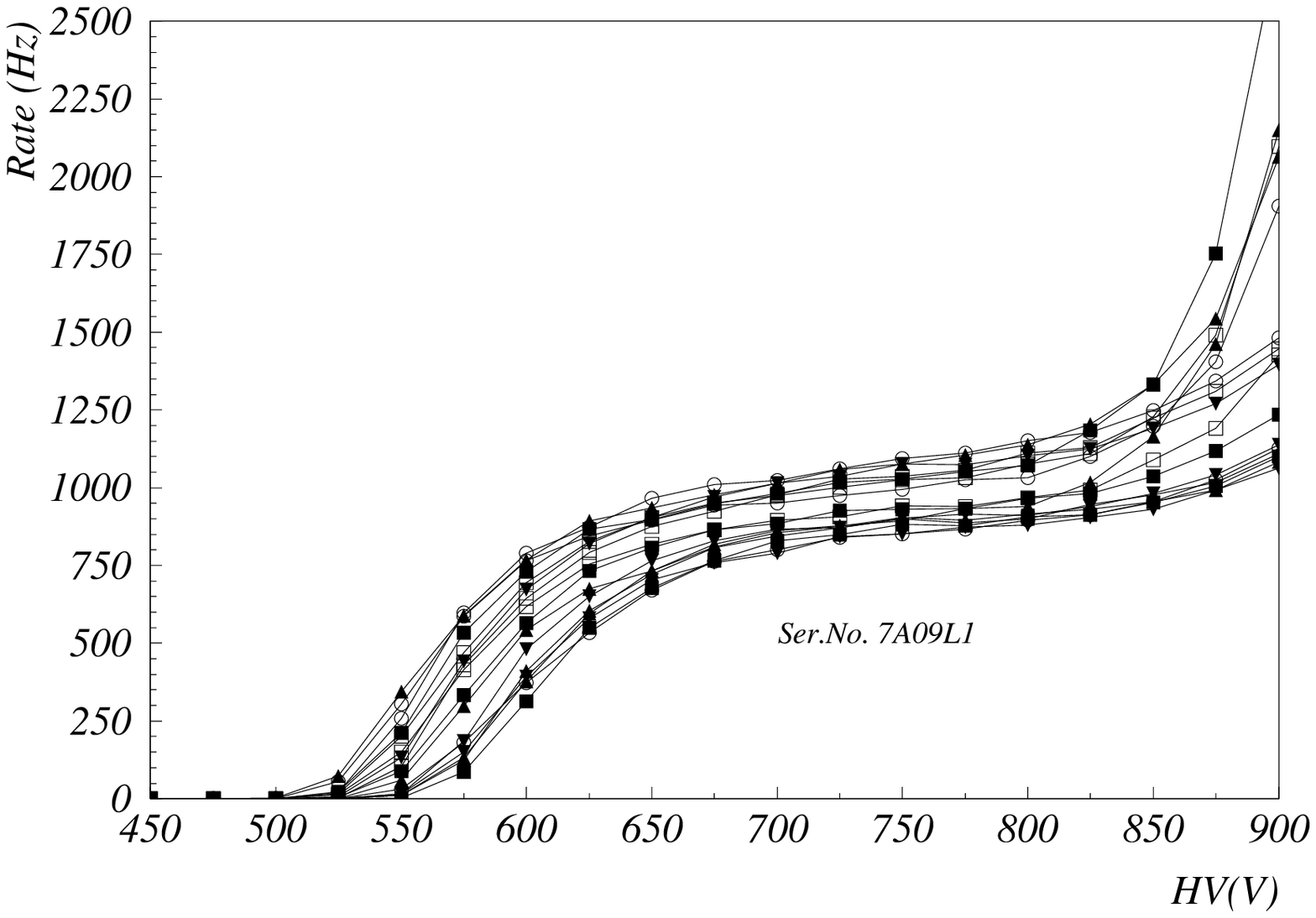,width=12cm}}

\vspace{-1cm}

\centerline{\psfig{file=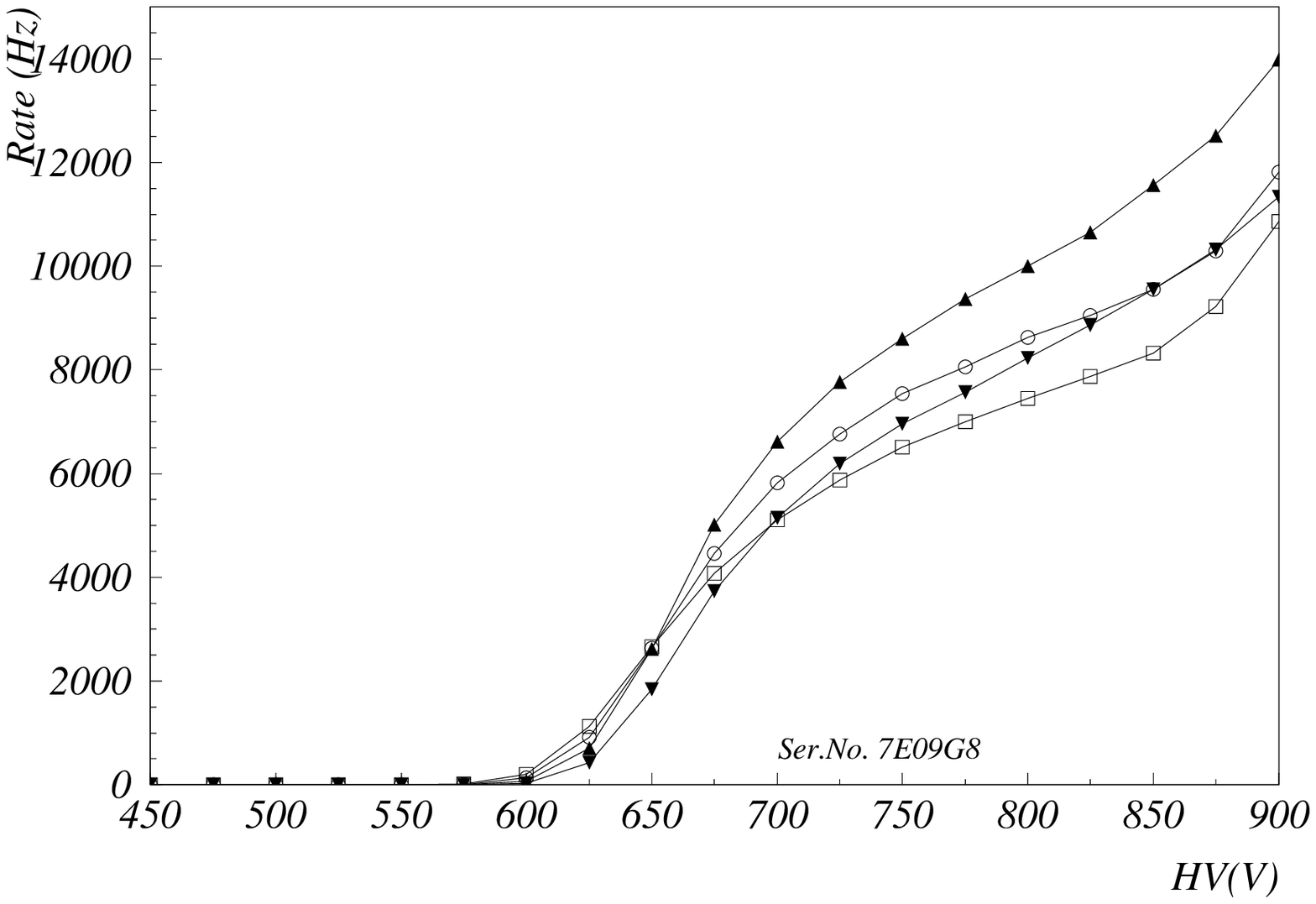,width=12cm}}
\caption[kk]{ The plateau curves i.e., the
dependence of measured rate as a function of the
high voltage for all channels of a representative M16 (above) and M4
(below) photomultiplier tube.
}
\label{scan}
\end{figure}

\begin{figure}[htb]
\centerline{\psfig{file=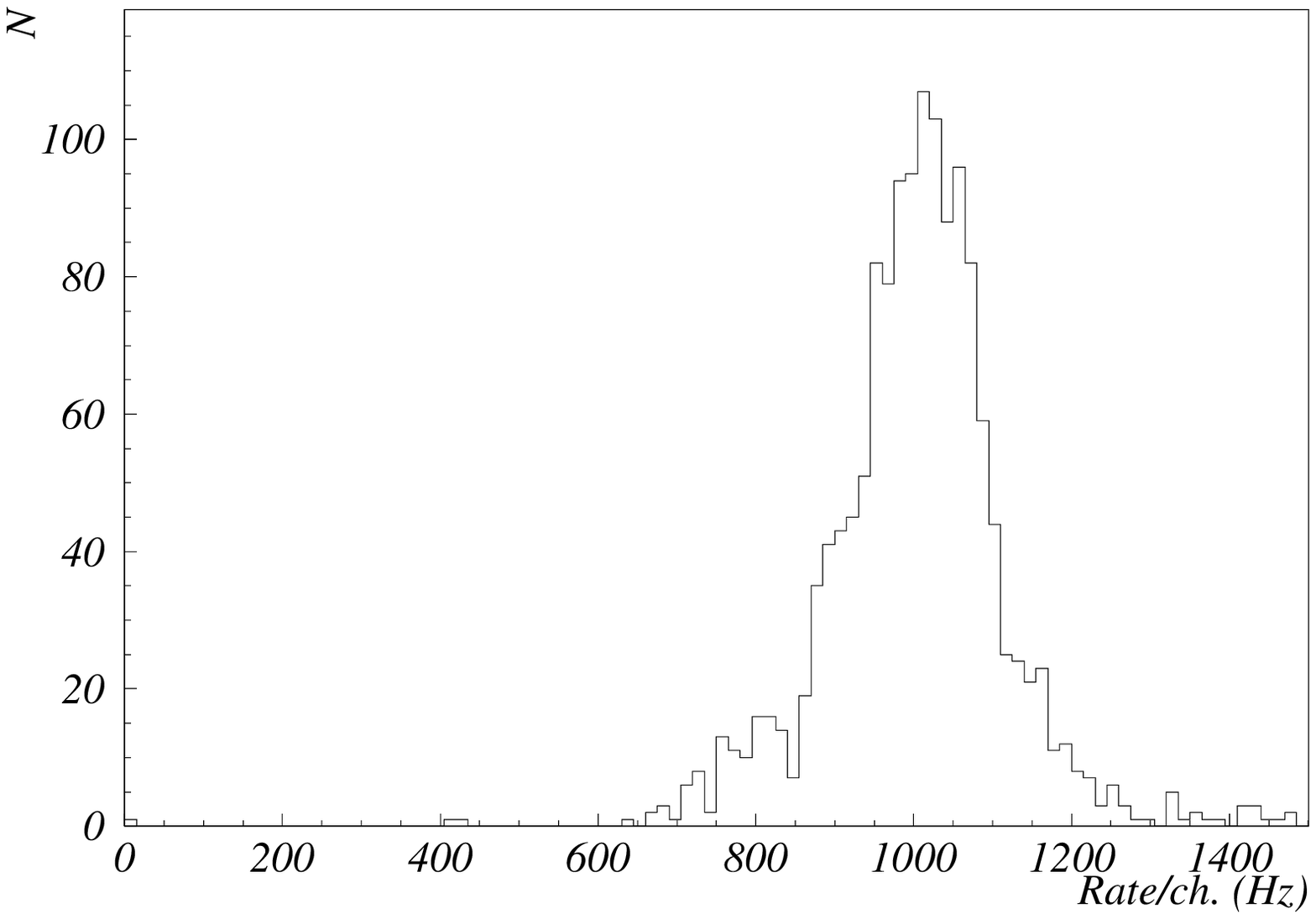,width=12cm}}

\vspace{-1cm}

\centerline{\psfig{file=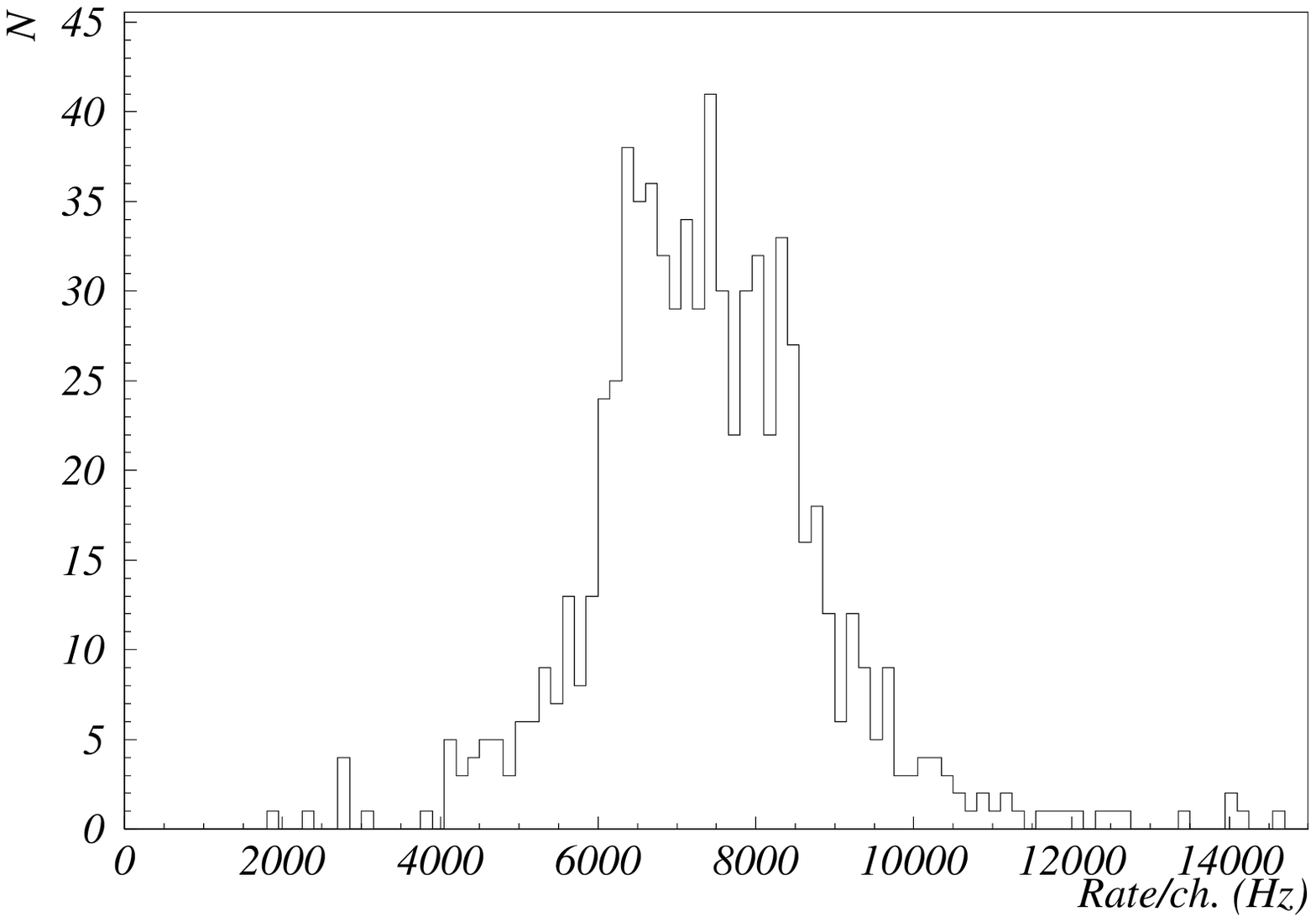,width=12cm}}
\caption[kk]{
Distribution of the number of PMT channels depending on the
rate per channel for the M16 (top) and M4 (bottom) PMTs
at the optimum high voltage, as determined in the quality assessment tests.
}
\label{hvset}
\end{figure}

The actual photon detector consists of  Hamamatsu 
multi-anode R5900-00-M16 and R5900-03-M4 photomultiplier tubes (PMTs).
In what follows they are denoted by M16 and M4.
The M16 version has 16 pads of 4.5~x~4.5~mm$^2$ each, with a
12-stage, metal-foil dynode system \cite{hamamatsu-cat}. 
The M4 version has 4 pads of 9~x~9~mm$^2$ each, and  10 dynodes. 
The quantum efficiency of the M16 photocathode
with borosilicate window 
has a broad plateau 
in the wavelength region between 300 nm to 500 nm
and a maximum value about 20\%. 
The M4 tubes
have a UV transparent window, which 
shifts the low wavelength cut-off to about 250~nm. 
The other PMT characteristics, such as the required cathode 
high voltage ($\le$ 1000 V),
current amplification (10$^7$), dark current ($\sim$ 1 nA), 
pulse rise time (0.8 ns), transit time spread (0.3 ns) 
are also satisfactory \cite{hamamatsu-cat}.
In the initial set of on-the-bench 
measurements \cite{pmt-first}, the single photon counting 
properties of the photomultiplier tube 
were investigated, in particular the efficiency for
single photon detection as well as the background count rate.
It was established that the PMT allows for  a
good single photoelectron detection efficiency (above $98\%$),
with very small cross talk (below $0.2\%$),
low background rate (few Hz per channel, Fig.~\ref{figbg}),
and acceptable
uniformity as shown in
Figs.~\ref{scan} and~\ref{hvset}
\cite{icfabul}.

\begin{figure}[bht]
\centerline{\epsfig{file=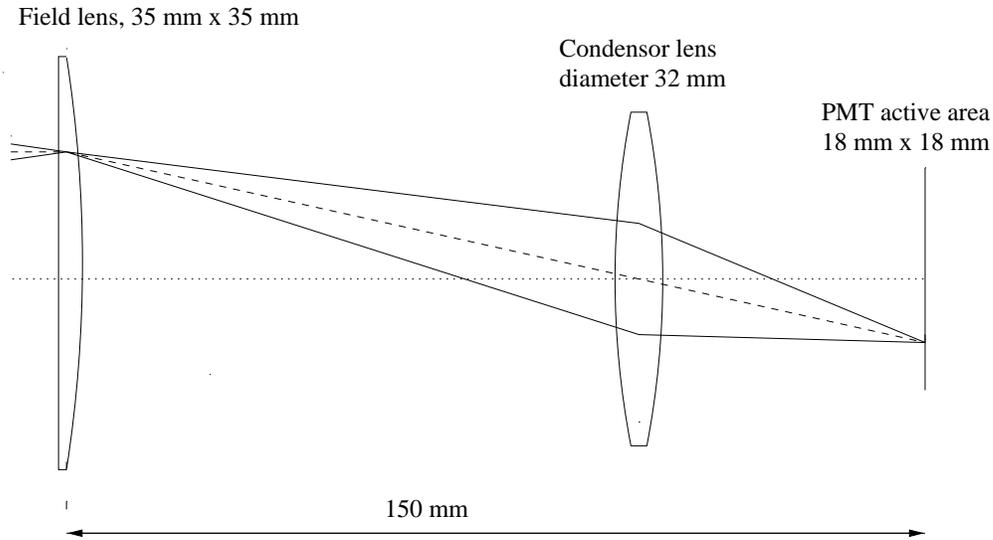,width=7.cm,
angle=-90}}
\caption[kk]{ The optical system for light collection and demagnification.
The two rays shown in full line correspond to photons with incident angles of
$\pm 100$~mrad
with respect to the dashed line (normal incidence).}
\label{lens_syst}
\end{figure}
\begin{figure}[bht]
\centerline{\epsfig{file=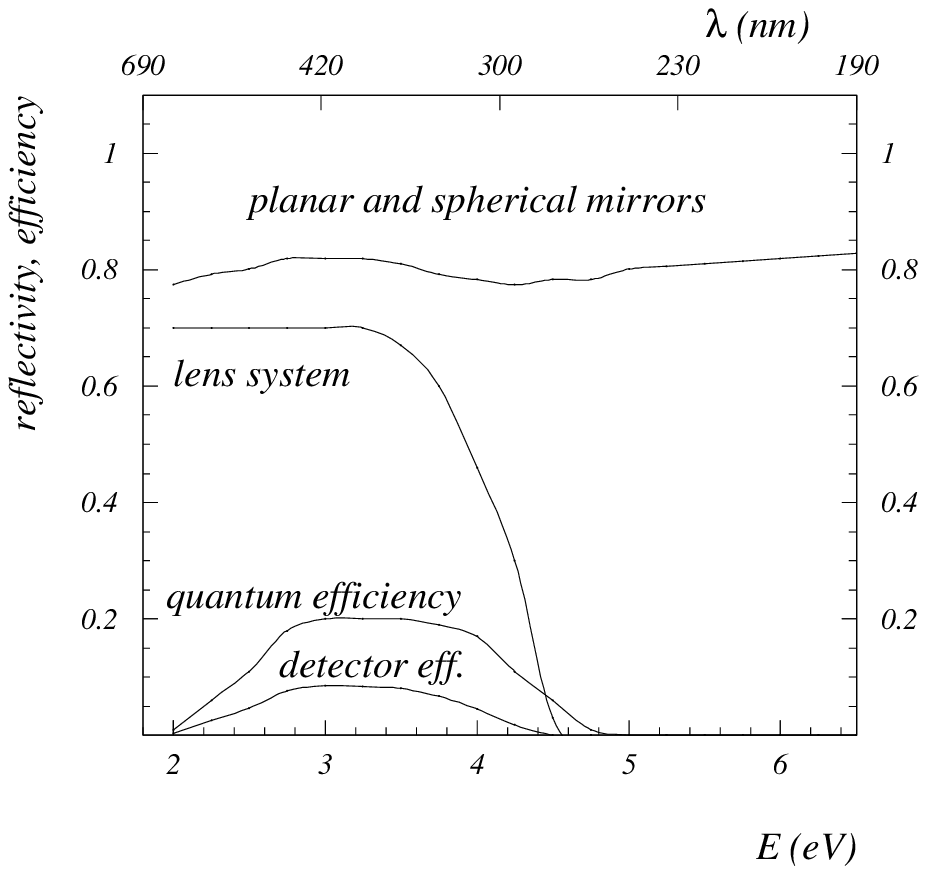,width=8.cm,
}}
\caption[kk]{Transmission of the lens system, reflectivities of
the mirrors,
quantum efficiency and overall detector efficiency
of the  photomultiplier tubes
as a function of photon energy and wavelength.}
\label{rich_qetr}
\end{figure}
The PMTs have the outer dimension
of $28\times 28$~mm$^2$ with an active area
of $18\times 18$~mm$^2$. They are positioned on a $36\times 36$~mm$^2$
grid. To increase the fraction of the active area, 
a two lens demagnification system (2:1) was designed, as shown in  
Fig.~\ref{lens_syst}
 \cite{lens-design}. 
The lenses are made  of UVT perspex~\cite{uvt}
with high
transparency over most of the wavelength region where the photocathode is
sensitive (Fig.~\ref{rich_qetr}). 
The angular acceptance of the optical system is also
satisfactory and  is uniform for incident angles below about 110 mrad 
\cite{michael-lens-test}.
The increase in the photodetector active area is achieved
at the expense of increasing the pixel size from the
4.5~x~4.5~mm$^2$ PMT pad size to 9~x~9~mm$^2$ pixel in the
central detector region.
The resulting angular measurement error (0.46 mrad)
slightly exceeds the spread due to 
dispersion (0.33 mrad).
For the outer detector region, with lower track densities and lower
typical track momentum,
the photon flux is smaller, and
the resolution requirements looser.
In order to reduce the detector cost, this detector part 
uses the M4 version of the tube with two times larger 
pads  (9~x~9~mm$^2$). 
The same lens system
thus results in
 18~x~18~mm$^2$ pixels which correspond to an angular measurement error
of 0.92~mrad.

\begin{figure}[bht]
\centerline{\epsfig{file=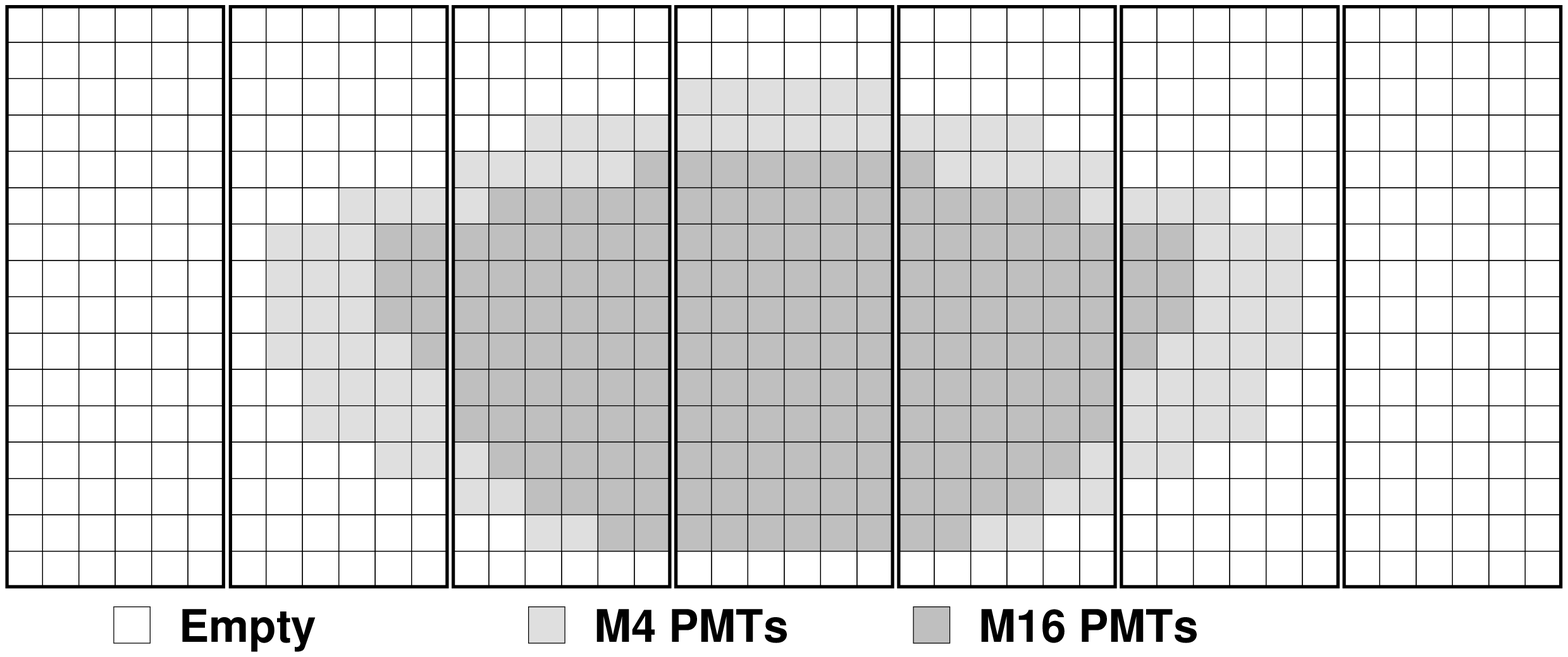,height=5.cm,}}
\caption[kk]{Arrangement of M4 and M16 type basic units 
         on the 7 supermodules of the upper focal plane.
Note that the left-most and the right-most  supermodules
are not instrumented.
}
\label{photdet-tiling}
\end{figure}

In order to reduce the contribution of spherical aberration to the overall 
resolution of the \v Cerenkov angle, an optimal surface of the \v Cerenkov 
photon detector was calculated \cite{det-form,roy-det-form}. 
Each half-detector (upper and lower)  
consists of 7 flat supermodules placed
to approximate the optimal 
surface, which is close to the shape of a flattened (ellipsoidal) cylinder. 
Such an arrangement also ensures better acceptance for the \v Cerenkov 
photons, which should be incident onto the flat supermodules at angles below 
110 mrad.
The supermodules are  rectangular 1.1
$\times$ 0.4 m$^2$ boxes containing a grid made from 1~mm thick soft
iron sheets that serve as magnetic shield and mounting structure for
the PMT base-boards and the light collection system. The two types of
photomultipliers were arranged according to occupancy and
reconstruction requirements as indicated in Fig. \ref{photdet-tiling}.
Altogether 1488 M16 PMTs and 752 M4 PMTs are needed to cover the
detector surface, thus totaling 26816 read-out channels.
%
%
%

Prior to the installation in the photon detector
tests of all the 2305 acquired PMTs (1543 M16's and 762 M4's) were
made in the laboratory \cite{beaune99}.
\v Cerenkov photons produced in a 1~cm thick quartz window
by  $\beta$ particles from a $^{90}$Sr
source were used as a stable light source \cite{sr-90}. 
Four PMTs were tested simultaneously, of which one was
used as a reference. For each photomultiplier tube, the source and 
background rates were recorded as a function of high voltage and threshold
setting. On the basis of these tests, the photomultipliers were
grouped according to similar high voltage characteristics, allowing all
PMTs within a group to be connected to the same high
voltage, thus minimizing the
number of
independent HV channels \cite{hv-report}. 
The results obtained for the optimal high voltage as
well as for the relative PMT sensitivity were compared to values provided 
by the manufacturer, and  good agreement was found \cite{beaune99}. 
The count rates of the reference  M4 and M16 photomultipliers,  recorded
during the quality assessment tests and  also continually measured
 after the tests, were used
in order to obtain an estimate of the long term
stability. The count rate decrease in two years
of operation is consistent with the 
known decay rate of the $^{90}$Sr source.

\subsection{Read-out and monitoring}

\begin{figure}[bht]
\centerline{\epsfig{file=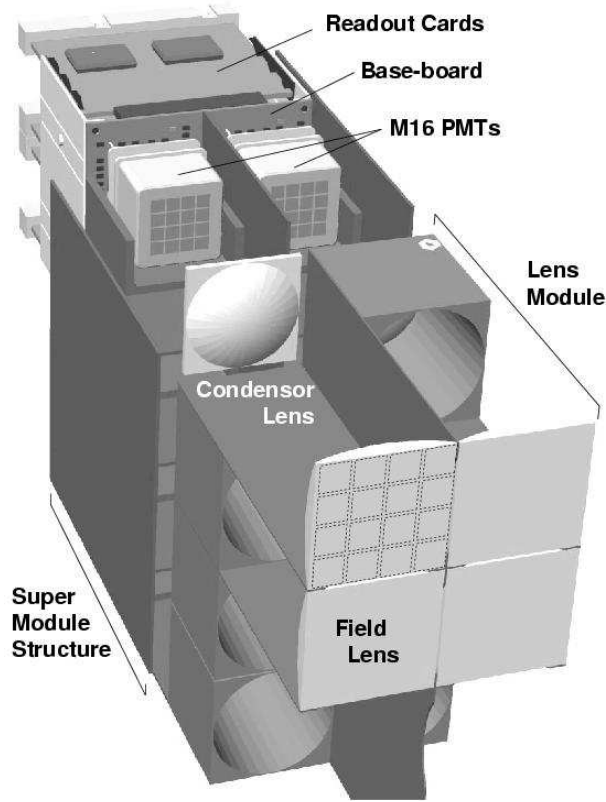,width=8.cm,
}}
\caption[kk]{Scheme of part of a supermodule of
the RICH photon detector.  }
\label{photdet_schema}
\end{figure}

To mount, power, and read out the photomultipliers,
two types of  base-boards were developed, one for M4 and one
for M16 PMTs. Each
base-board is a light-tight, four-layer (M16) or two-layer (M4)
circuit-card, 70 $\times$
70 mm$^2$ in size, equipped with surface mount components.
One side of the board
holds four custom designed sockets and
voltage divider chains, appropriate for
the particular type of PMT;
the other side contains one (M4) or four
(M16) connectors for the 16-channel readout-cards, attenuation
circuitry, and two daisy-chainable high voltage connectors.
The voltage divider chain is composed of a sequence of
1~M$\Omega$ and
560~k$\Omega$
resistors that
was optimized for single photon counting at high rates~\cite{asd8-dp}.
The
base-boards are mounted in such a way that the PMTs are positioned
at  centers of the $36 \times 36$~mm$^2$ grid
(see Fig. \ref{photdet_schema}).

%


Between two
and five base-boards, with their PMTs selected for similar optimal
operating voltage (ranging from  750~V to 890~V),
are powered by a single output
of a 160-channel CAEN SY527 high-voltage supply \cite{hv-report}.

The front end electronics board employed in the system is a 
 16-channel board~\cite{pohl}
based on the ASD8,
an amplifier, shaper, and
 discriminator chip~\cite{asd8}.
The board 
was developed for the Outer Tracker (OTR),
the main tracking system of the HERA-B spectrometer.
Since the photomultiplier tube signals are typically considerably
higher than the pulses for drift chambers, a charge divider was added
in front of the ASD8 board (1:10 for M16, and 1:5 for M4 PMTs).
The front end electronics  boards
were arranged  according to their optimal  threshold voltage
by using test data which were obtained
by varying pulse height and threshold voltages \cite{igna}.
The same test was also used to reject faulty boards.

The low voltage supply system for the ASD8 boards 
is  powered by four KEPCO ATE 6-100M supplies,  
delivering  voltages of ${\rm +3\ V}$ and   ${\rm -3\ V}$ separately
to the upper and lower photon detector halves \cite{ds-dr}.    
Analog outputs 
of  16  photon detector  channels are 
transmitted to the electronics hut
for diagnostic purposes.


  The signals from the ASD8 cards are transmitted over 7.5 meter long 16
channel twisted pair cables to the Front End Drivers (FEDs).  For each FED
set there are 4 daughter cards with 16 cable connections, and one mother
card, such that each FED set can accomodate up to
1024 channels~\cite{fed-system,kendal-phd}.  The RICH
uses 28 such sets. The FED mother board provides the interface to the
HERA-B Fast Control System, which provides a clock signal for
synchronization as well as triggers to indicate event forwarding.  The
mother board also provides the interface to the HERA-B DAQ System via the
SHARC protocol \cite{sharc}, as well as a ring
buffer memory capable of storing 128
events, pending the First Level Trigger decision.

%


A sizable fraction of the operating parameters of the RICH counter 
(low voltage and threshold for the front end electronics boards, 
temperatures at the photon detector and power supplies) are
controlled by a Field Point \cite{fieldpoint} based system. Two 
separate systems are used for monitoring  the high voltage
\cite{hv-report}
and steering  the gas system \cite{gas-system}. The overall control
of these systems is implemented with a server-user interface system. The
server process is running on a VME computer connected to the Field Point
system through a RS232 cable. In the case of a malfunction
(if the temperature,
threshold, or low voltage are out of range) the server switches off the
low voltage power supplies. The parameters can be monitored and set
via a 
tcl/tk and BLT based graphical user interface program running on
a linux PC and communicating with the server through
a TCP-IP based protocol.
The RICH low voltage and high voltage slow control systems are integral parts
of the overall HERA-B slow control system \cite{slow-control}.

As a triggerable light source for testing the photon detector in
the periods without beam induced 
reactions, a system of light emitting diodes (LED) was installed at
the sides of the photon detector windows \cite{led-system}. The system
provides good illumination over the whole photon detector area.


\subsection{Expected performance}

The number of detected photons on a ring measured by a \v Cerenkov detector is
often parametrized as $N = N_0 L \sin^2 \theta _c$, where $L=2.82$~m
is the length of the
radiator, $\theta_c$ is \v Cerenkov angle,
and $N_0$ is the detector response parameter.
From the data available on the quantum efficiency \cite{hamamatsu-cat}, 
mirror reflectivities \cite{mirror-tilling}, 
window and optical system transmissions \cite{lens-design}, 
one calculates the merit factor $N_0 = 41$~cm$^{-1}$.
Accordingly,
the expected number of photons for particles approaching the velocity 
of light ($\theta_c=52.4$~mrad) amounts to 32.

Individual
contributions to the expected single photon resolution are summarized
in Table~\ref{tabres}.
The optical error includes contributions from spherical aberration,
mirror quality and alignment of mirrors, while photon detector
resolution is dominated by detector granularity with a small contribution
of the alignment of photon detector components.
The contribution of multiple scattering in the RICH counter~\cite{glassel}
depends on momentum
and is given
by $\frac {3.5\;\rm{mrad}}{p\;\rm{(GeV/c)}}$.
The resulting single photon resolution,
$0.65\;\rm{mrad} \oplus \frac {3.5\;\rm{mrad}}{p\;\rm{(GeV/c)}}$
($1.02\;\rm{mrad} \oplus \frac {3.5\;\rm{mrad}}{p\;\rm{(GeV/c)}}$)
for the regions covered by M16 (M14) PMTs, does not include
the contribution from the track direction as given by
the tracking system, and will therefore be
referred to as intrinsic resolution.

\begin{table}[htbp]
\caption[kk]{
Contributions to the single photon angular resolution for M4 and M16
photomultiplier tubes.
}
\begin{center}
\begin{tabular}{|c||c|c|}                                                   
\hline
  & &  \\ 
error source & M16 PMT &
M4 PMT \\
  & &  \\                                                                     
\hline                                                                    
\hline
photon detector resolution & 0.50~mrad & 0.93~mrad
\\ \hline
   $C_4F_{10}$ dispersion  &    
 \multicolumn{2}{c|} {0.33~mrad}                 
\\ \hline
optical errors  &
\multicolumn{2}{c|}
{0.25~mrad}
\\ \hline
multiple scattering &
\multicolumn{2}{c|}
{$\frac {3.5\;\rm{mrad}}{p\;\rm{(GeV/c)}}$}
\\ \hline
total intrinsic resolution &
$0.65\;\rm{mrad} \oplus \frac {3.5\;\rm{mrad}}{p\;\rm{(GeV/c)}}$&
$1.02\;\rm{mrad} \oplus \frac {3.5\;\rm{mrad}}{p\;\rm{(GeV/c)}}$
\\ \hline
%
\end{tabular}                                                               
\end{center}                                                                   
\label {tabres}    
\end{table}

\section{Data Analysis}

Several computer programs were developed for the analysis and
monitoring of the data collected by the RICH counter.
In the initial phase of the commissioning
of the HERA-B spectrometer, when major parts of the main tracking system 
were not yet available,
a stand-alone ring search algorithm was employed
\cite{rise,denis-phd}.
The program searches for rings with a given
ring radius,
starting at the maximum value and scanning down to $70\%$
of the maximal ring radius.
Only rings with a sufficiently high signal over 
background ratio are stored, and the corresponding hits
are excluded from further ring search.
As described in section 6,
the reconstructed  ring center gives  information that
can be used in tracking.
The algorithm is routinely used to monitor the
current maximal value 
of the \v Cerenkov angle \cite{thc-monitor}
 which is, in turn, used as the input for all particle identification 
algorithms.
Small variations of refractive index and thus of the \v Cerenkov
angle are  caused by pressure and temperature variations
of the radiator gas.
We note that several alternative  ring search algorithms
were also tested on the data  \cite{ringsearch2}.

For highly populated events, typical for the reactions recorded in the
HERA-B spectrometer, additional information helps to
improve the particle identification capabilities.
The information from the tracking system, track direction and momentum,
is used to 
determine the expected position of the ring center and the
expected ring radius for various 
particle hypotheses. To evaluate the likelihood for each
individual hypothesis, extended 
maximum likelihood methods \cite{baillon} are applied, either directly
\cite{rire,arinyo-phd} or combined with an expectation-maximization algorithm
\cite{riter,rok-phd}. The latter is
described in more detail in section 5. 

Another way to take the track information into account is to perform
a seeded ring search in regions given by the track direction and
momentum \cite{denis-phd}.
Since only restricted regions corresponding to five hypotheses
are scanned, this approach is quicker than the stand-alone
ring search.
Recently, a new particle identification method was developed
which maximizes the 
likelihood value by allowing the track direction to vary
within errors given by the track fit 
\cite{roy-likelihood}.

\section{Measurements and Results}

\subsection{Photon detector performance}

\begin{figure}[t]
\centerline{\psfig{file=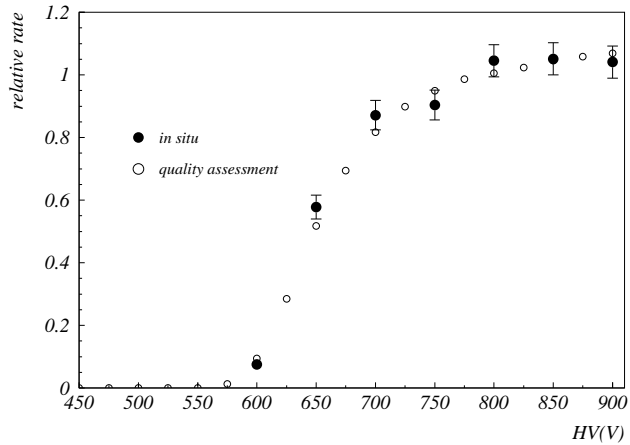,width=10.cm,clip=}}
\caption[kk]{ The plateau curve for a typical M16 tube. 
The data labeled 'in situ' represent values measured during the photon 
detector testing in its
final position, while the curves labeled  'quality assessment' correspond to 
measurements made in the laboratory during on-the-bench quality tests.
The two sets of 
data are normalized to the average of the three points at 750~V, 800~V and 
850~V. }
\label{hvscan}
\end{figure}
\begin{figure}[hbt]
\centerline{\epsfig{file=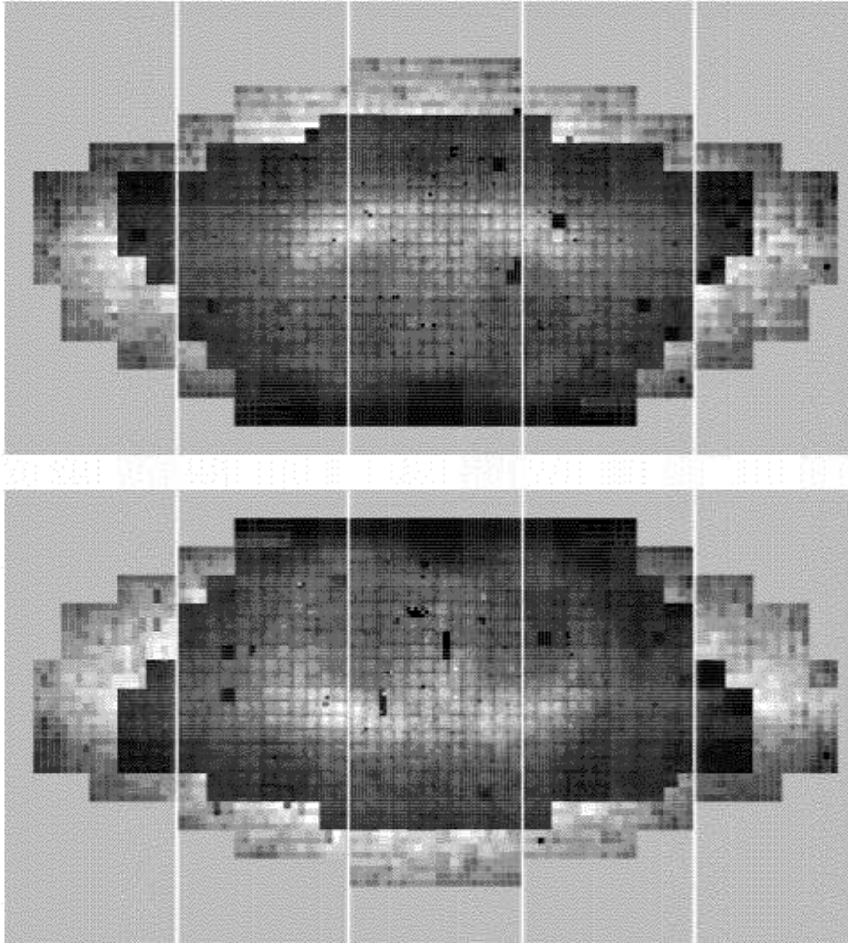,height=12.cm,
angle=270,clip=}}
\caption[kk]{ The occupancy of the upper and lower photon detectors
shows the region occupied by M16 (inner region) and M4 PMTs (outer region).
The peak values, shown in white, correspond to rates 
around 1.5~MHz per channel.
}
\label{phdet-occu}
\end{figure}
\begin{figure}[hbt]
\centerline{\epsfig{file=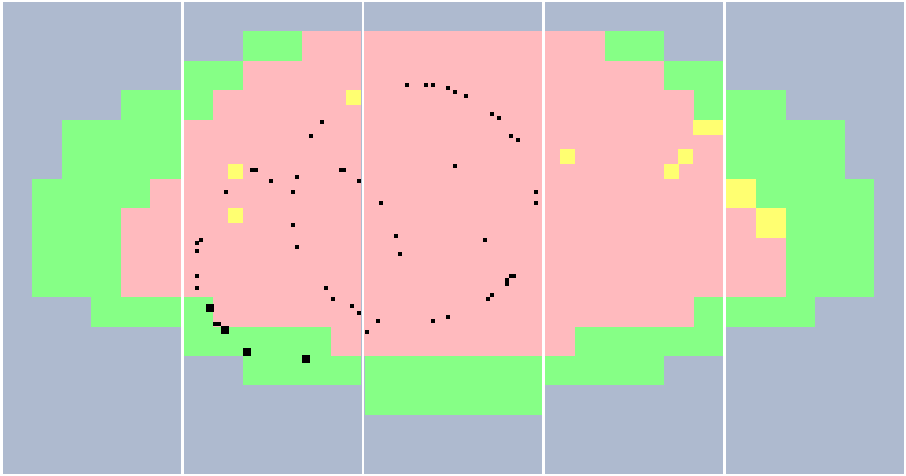,height=12.cm,
angle=270,clip=}}
\caption[kk]{ Event with two \v Cerenkov rings on the lower photon detector.}
\label{two-rings}
\end{figure}

%
In the first step of the commissioning of the system,
some basic parameters were investigated.
Fig.~\ref{hvscan}
gives the count rate versus high voltage for 
a representative  
photomultiplier. It is shown that the curves measured in-situ with the HERA
proton beam agree nicely with the $^{90}$Sr source measurements. The
time averaged
occupancy is shown in Fig.~\ref{phdet-occu}, 
where the region occupied by M16 PMTs is clearly 
distinguished from the region occupied by M4 PMTs.
The fraction of dead channels is about $2\%$, and is partly due to
missing PMTs and partly due to 
PMTs which were damaged during the installation phase.
Only about 0.3\% of the 26816 channels were 
found to be noisy, and were excluded from the analysis.
As a result, very clean rings can be observed in low multiplicity events
such as the one shown in Fig. \ref{two-rings}. The absence of random hits
in the event also confirms the earlier result that the  background due to 
scintillation in the radiator gas is negligible.
%

\subsection{Number of photons}

\begin{figure}[hbt]
\centerline{\epsfig{file=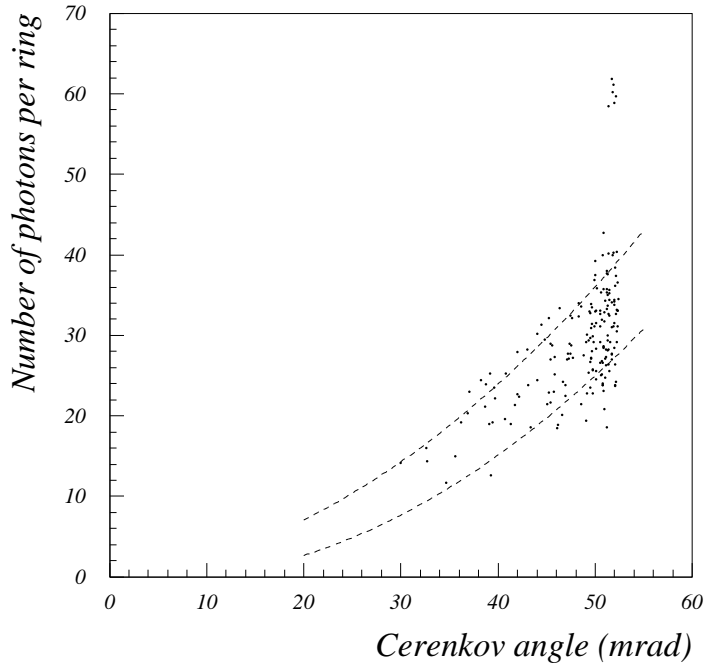,height=10.cm}}
\caption[kk]{ Number of photons on a ring versus the corresponding \v Cerenkov angle.
The points represent measurements of individual rings, while the dashed lines
correspond to one sigma deviations from the predicted dependence. The small group of events at
$\theta _c$ = 52 mrad, which has about two times the number of photons per
ring, probably corresponds to  overlapping $e^+ e^-$ pairs.
}
\label{nph-thc}
\end{figure}

\begin{figure}[hbt]
\centerline{\epsfig{file=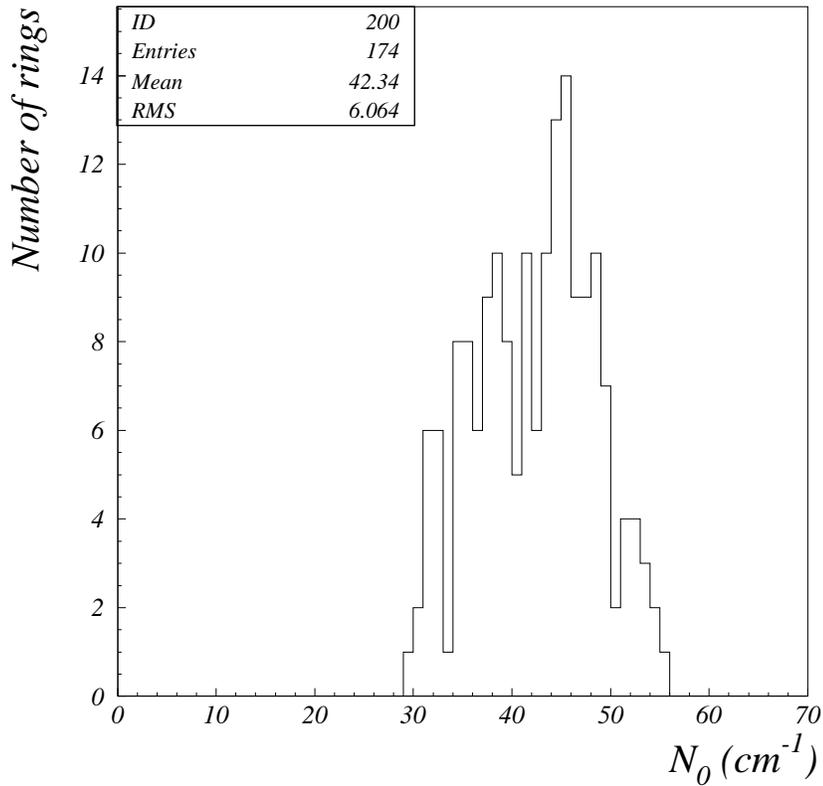,height=12.cm,}}
\caption[kk]{Distribution of the RICH response parameter $N_0$
obtained from measured individual \v Cerenkov rings (see text).}
\label{rich-n0}
\end{figure}

In order to obtain the basic RICH parameters from the data, 
 initially only events with a small number of \v Cerenkov
rings were analyzed  
\cite{ds-dr}. For each ring, the number of photon hits
was counted and the
\v Cerenkov angle was determined, obtaining a point in
Fig.~\ref{nph-thc}. It is seen that the observed rings follow
the expected
dependence of the number of photons versus the \v Cerenkov angle. The largest
measured \v Cerenkov angle of 52.4 mrad also corresponds to the expectation for
$\beta = 1$ particles ($\cos\theta _c = 1/n$).
The parameter $N_0$ was
calculated for
each of the isolated rings and the distribution is plotted in
Fig.~\ref{rich-n0}, from which it is seen that the mean value is 42 cm$^{-1}$.
This value corresponds to 33 detected \v Cerenkov photons for
$\beta=1$ particles.

%

\subsection{Angular resolution}

\begin{figure}[hbt]
\centerline{\epsfig{file=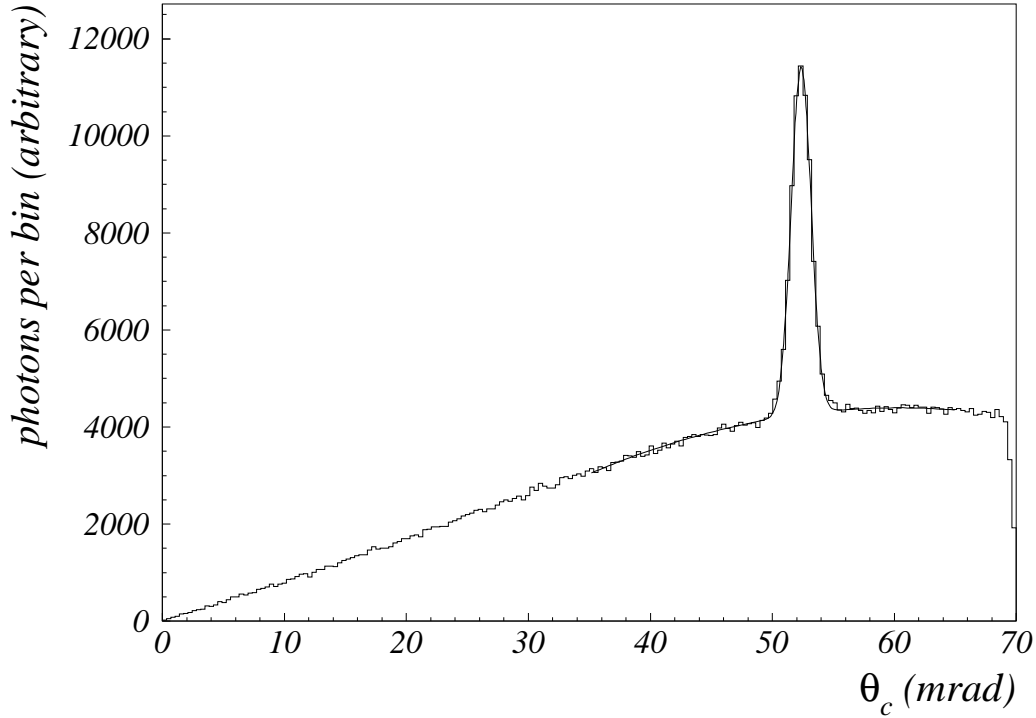,height=10.cm,}}
\caption[kk]{Single photon resolution for M16 PMTs: distribution of
the measured \v Cerenkov angle for individual hits in the case of muons
(from $J/\Psi$ decays)
with momenta above 40 GeV/c. The background below the peak comes
from wrong track-photon combinations.
A fit with a gaussian function and a second order polynomial gives
$\sigma=0.81$~mrad.
}
\label{thc-singph-hist}
\end{figure}
\begin{figure}[hbt]
\centerline{\epsfig{file=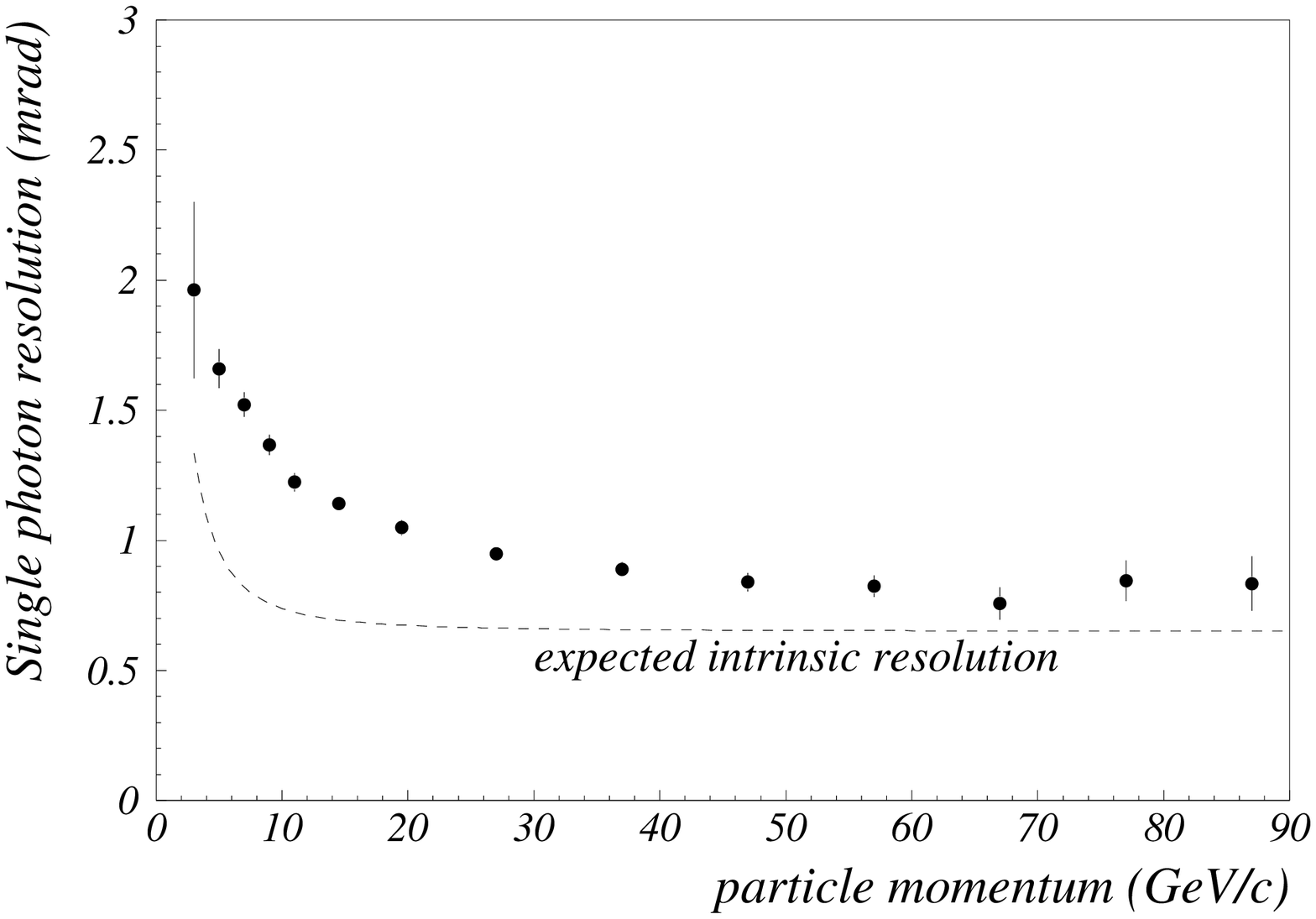,height=10.cm,}}
\caption[kk]{The measured single  photon \v Cerenkov  angle resolution  
versus particle momentum for the region covered by M16 photomultiplier 
tubes.}
\label{fig:resolut}
\end{figure}
To determine the  \v Cerenkov angle uncertainty given by a 
single \v Cerenkov photon, tracks with known identity are selected.
For each track 
within a given momentum interval, one fills  a
histogram with the photon hits at their reconstructed \v Cerenkov angle, so 
that the hits due to \v Cerenkov rings should show up as a peak, as shown on
Fig.~\ref{thc-singph-hist}.
The distribution was fitted with a
Gaussian function and polynomial background.
The resulting momentum dependence of the single photon resolution
is shown in Fig.~\ref{fig:resolut}. 
As expected, the single photon  resolution depends on the
particle momentum.
For the high momentum tracks the resolutions
of  0.81~mrad and 1.0~mrad were measured for the regions
covered by M16  and  M4 PMTs, respectively.
The difference between the measured single photon resolution
and the expected intrinsic resolution  (as shown in Fig.~\ref{fig:resolut})
is attributed to the error in the
measurement of charged particle trajectory, which was not taken
into account in the calculation of the expected intrinsic \v Cerenkov angle
uncertainty (Table~\ref{tabres}).

 The intrinsic resolution of the RICH counter was  
 studied by the stand-alone ring search analysis 
\cite{richstat-roy}, and  turned out to be  consistent
with the expected values.
The intrinsic resolution in \v Cerenkov angle given
by a single \v Cerenkov photon was 
also  obtained with the aid of hits in the electromagnetic
calorimeter \cite{imi2000}.
For the selected set of tracks
with cluster energy above 50 GeV,
the center of the \v Cerenkov ring is accurately
determined from the straight line
connecting the center of the cluster with the wire target  (the
magnetic field was off for this measurement).
The single \v Cerenkov photon resolutions obtained
correspond to $\sigma$ = 0.7 mrad 
($\sigma$ = 1.0 mrad) for the M16 (M4) PMTs,
in good agreement with expected values given in Table~\ref{tabres}.

The  particle  identification
capabilities are affected  by  the high multiplicity of rings in a
single event. Typically 50 overlapping  rings are present in an event
of which  about 1/3 could be  associated with   measured tracks,
while the rest belong to tracks coming from secondary interactions.
As a result,
channel occupancies in some regions are as large as 25\%.
We note that in such an environment the resolution in the \v Cerenkov
angle measurement of a track does not simply scale with the inverse
of the square root of the number of detected photons.

\section{Particle Identification Performance}

\begin{figure}[hbt]
\centerline{\epsfig{file=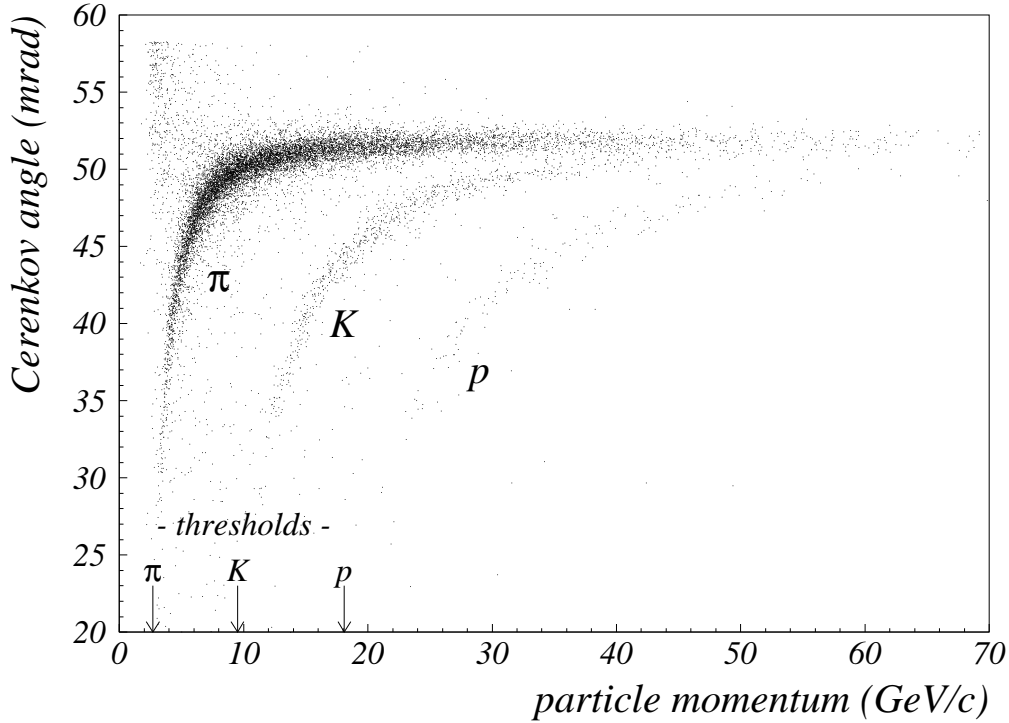,height=10.cm,}}
\caption[kk]{Reconstructed \v Cerenkov angle versus particle
momentum.  Bands for  pions, kaons  and protons  are  clearly visible.
}
\label{fig:cerplot}
\end{figure}

To illustrate the particle identification capabilities,
a maximum likelihood fit
was performed on \v Cerenkov angle distributions for individual tracks. 
Fig. \ref{fig:cerplot} shows the resulting bands for
different particles in the plot
of  reconstructed \v  Cerenkov angle  versus momentum,  as  obtained by
analyzing a sample  of measured data.

Two methods are routinely used for particle identification, a ring search
based and a track based algorithm. The ring search based particle
identification combines the stand-alone and seeded approach
discussed in Section 3.5.
In the first step, the stand alone reconstruction is attempted
and rings found are matched with tracks. In the next
step, tracks found by the vertex detector and the tracking detectors,
that were not  matched with a RICH ring, are used as seeds in a seeded
ring search.

For the track based particle identification the extended
likelihood method is used~\cite{baillon},
combined with  the expectation-maximization algorithm. The method is
described in detail in  \cite{riter,rok-phd}.
The first step  in the method is the calculation  of \v Cerenkov angle
(as described in \cite{det-form}) 
for all  track-photon pairs  with the \v  Cerenkov angle  smaller than
70~mrad.  The  \v Cerenkov angles of  the pairs are stored  in a list,
together with  a probability  (weight, $w_i$) that  the photon  from a
pair is emitted by the track  of that pair.  Initially the weights are
set  to  $1/N_{track}$, where  $N_{track}$  is  the  number of  tracks
corresponding  to  a  specific  photon  hit.  In  the  next  step  the
expectation-maximization algorithm  is applied  to the list  of pairs.
The  result  of the  algorithm  is  a set  of  new  values of  weights
$w_i$. In  this way the track-photon  pair which is more  likely to be
the right one, gets a larger weight.

In the last step  the extended likelihood probabilities are calculated
for 6 possible hypotheses for the identity of a track: {\it electron},
{\it muon}, {\it pion}, {\it  kaon}, {\it proton} and {\it other}. The
last one  represents the  case, where the  distribution of  \v Cerenkov
angles of photons for a given track is consistent with the background.
The  resulting  likelihood  probabilities  have the  range  of  values
between 0 and 1 and are normalized so their sum equals to one.
We note 
that the background level is determined on an event-by-event basis 
for each individual track.

The selection  of tracks  belonging to a  particular particle  type is
made by applying a cut  on the appropriate likelihood.  For easier use,
three levels  of selection are defined:
{\it  soft}, {\it  medium} and
{\it hard}.  The  levels are  set to  0.05 ({\it  soft}),  0.50 ({\it
medium})  and 0.95  ({\it hard})  for  pion selection  and 0.05  ({\it
soft}), 0.30 ({\it  medium}) and 0.95 ({\it hard})  for kaon or proton
selection.


To determine  the efficiencies and  misidentification probabilities from
the measured data, an a priori knowledge of particle types is required. The
following reconstructed decays were used:
$(1)$  $K_S^0 \rightarrow \pi^+ \pi^-$, as a source of pions, 
$(2)$  $\Lambda \rightarrow p \pi^-$ and
$\bar{\Lambda} \rightarrow \bar{p} \pi^+$ as a source of protons,
antiprotons and pions, and
$(3)$ $\phi(1020) \rightarrow K^+ K^-$, as a source of kaons.
For the  first two decays  a very clear  signal in the  invariant mass
plot  is obtained  by cutting  on  the secondary  vertex distance  and
removing the reflections ($\Lambda$, $\bar{\Lambda}$ in $K_S^0$, and
similarly for $\Lambda$ and $\bar{\Lambda}$).
The $\phi$ decay products come
from the  primary vertex  together with
roughly  10  other  particles.   The invariant  mass  plot  shows  a  huge
combinatorial  background.   To  reduce  the  background,  one  of  the
particles was   used to tag the decay by identifying  it as a kaon.
The other  one was  used
for efficiency and misidentification evaluation.
In all  cases, the number  of particles of  a given type  surviving the
selection criteria, and the number of all particles of the same type in
a selected momentum bin were obtained by fitting a Gaussian plus a
linear function to the invariant mass plots.

\begin{figure}[hbt]
\centerline{\epsfig{file=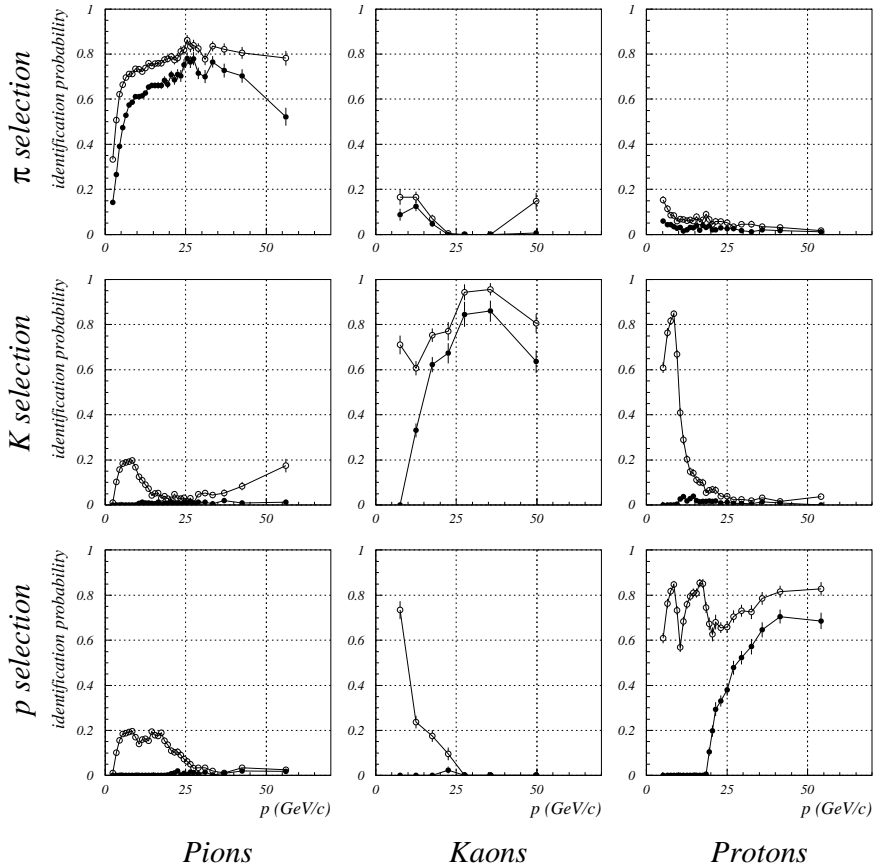,height=12.cm,}}
\caption[kk]{Momentum dependent identification probability plots for 
medium (open circles) and hard (full circles) 
selections. Plots on the diagonal show the efficiency,
plots
left or right to the diagonal show the mis-identification probability.
}
\label{fig:effi}
\end{figure}
The results  are presented in Fig.~\ref{fig:effi}.
At medium selection
criteria, kaon identification efficiency for momenta between
10 GeV/c and 60 GeV/c lies in a range from 60\% to 95\%,
with typically about 5\% pion mis-identification probability.
The corresponding proton identification efficiency for momenta
from 20 GeV/c to 60 GeV/c lies in a range between 60\% and 80\%
with less than 5\% pion mis-identification probability.
In the region below kaon and proton
 \v Cerenkov threshold,  the efficiencies are  70\% for
kaons (from 5 GeV/c to 10 GeV/c), 
and  from 60\% to
80\% for  protons (momentum region between 5 GeV/c and 20 GeV/c)
with less than 20\%  pion mis-identification.
Below  5~GeV/c the  performance  is degraded  due  to track  direction
uncertainty caused by multiple Coulomb scattering,
and a small number of \v Cerenkov photons for
pion tracks.

\begin{figure}[hbt]
\centerline{\epsfig{file=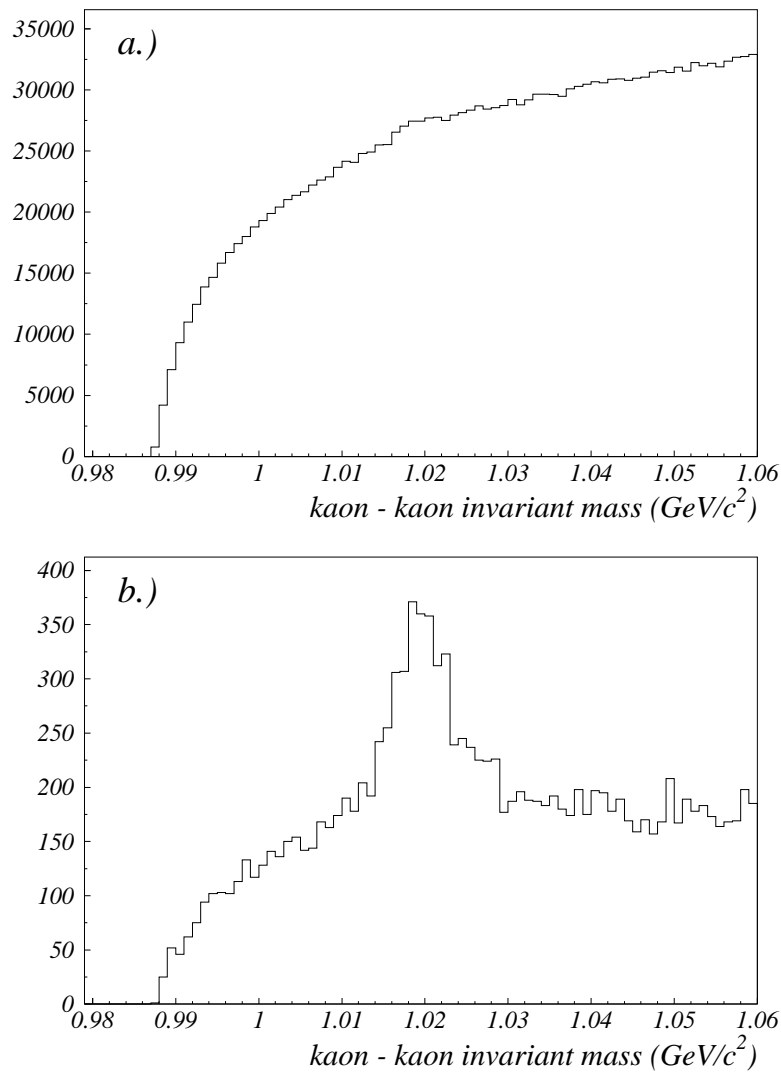,height=15.cm,}}
\caption[kk]{Improvement of signal to background for
invariant mass distribution of
$\phi \to K^+ K^-$:
before (a) and after applying hard selection on both kaons (b).
}
\label{phis}
\end{figure}

\begin{figure}[hbt]
\centerline{\epsfig{file=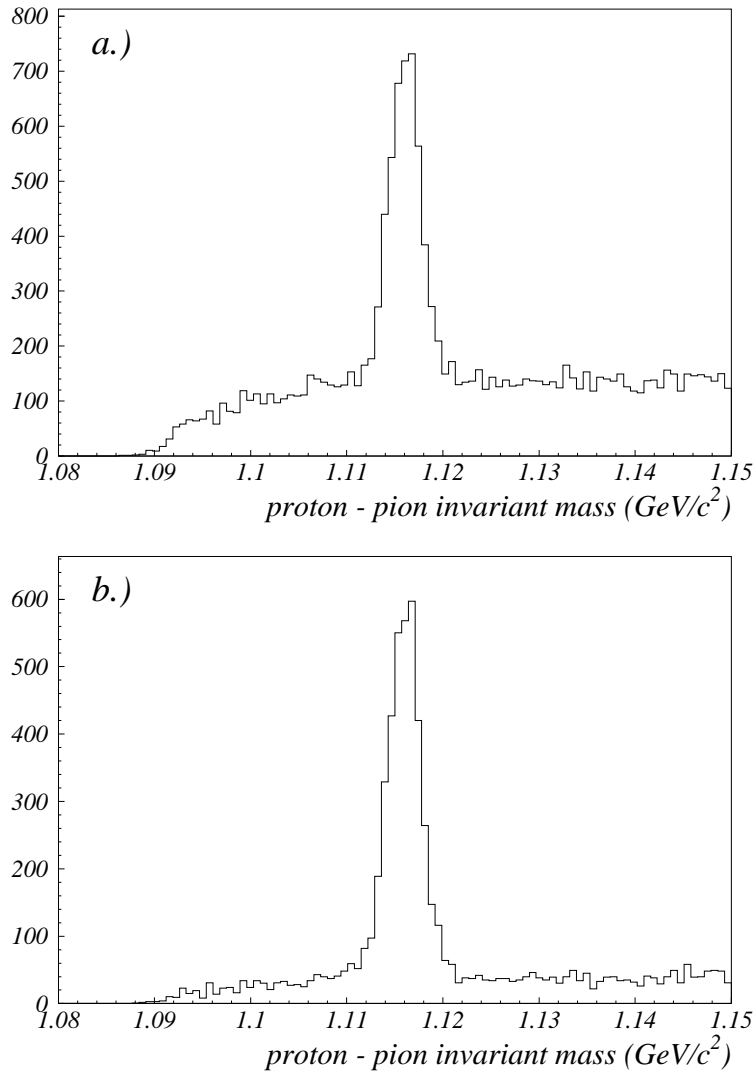,height=15.cm,}}
\caption[kk]{Improvement of signal to background
for invariant mass distribution of
$\Lambda \to \pi ^- p$ in the kinematic region  where the
$K_S^0\to\pi^+\pi^-$ hypothesis cannot be excluded:
a) no particle identification,
b) for the positively charged particle, the proton likelihood
was required to exceed the pion likelihood.
}
\label{lambdas}
\end{figure}

\begin{figure}[hbt]
\centerline{\epsfig{file=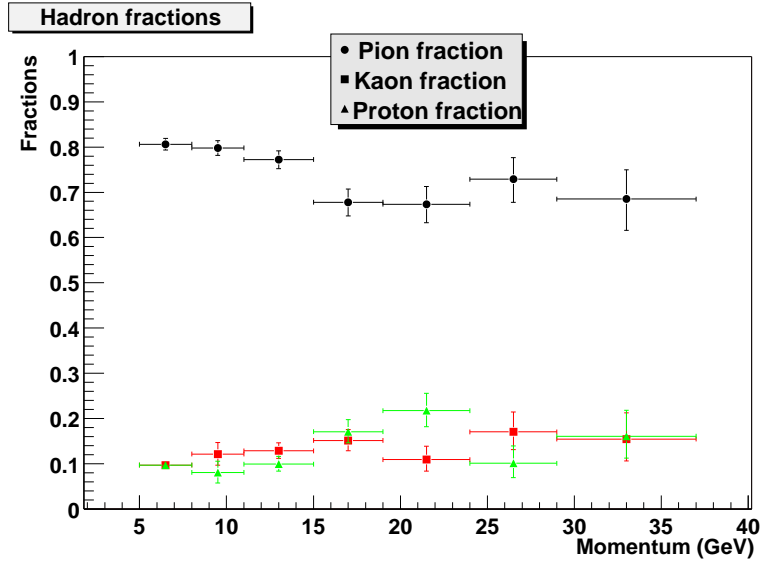,height=10.cm,angle=-90}}
\caption[kk]{Measured hadron fraction as a function of particle momentum.
}
\label{had-frac}
\end{figure}

Examples of performance and  impact on physics analyses are shown in 
Figs.~\ref{phis}-\ref{had-frac}.
From Figs.~\ref{phis} and~\ref{lambdas}, it is seen that
the inclusion of the requirement that both tracks are identified
improves the signal to background ratio in the
$\phi\to K^+K^-$ and $\Lambda \to \pi^- p$ decays.
By using the particle identification
as provided by the RICH counter, a measurement of hadron fractions
was performed~\cite{arinyo-phd}.
The results shown in Fig.~\ref{had-frac}, agree with
Monte Carlo predictions.


%

%

\section{RICH as a Tracking Device}

\begin{figure}[hbt]
\centerline{\epsfig{file=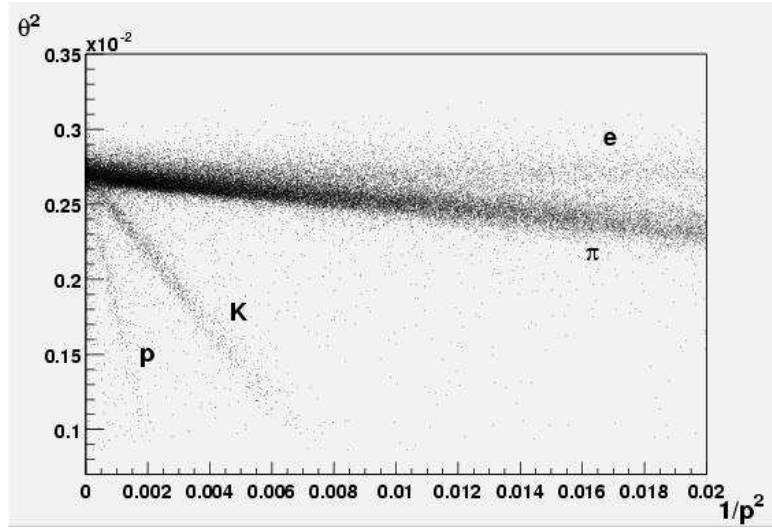,height=7.cm,}}
\caption[kk]{$\theta ^2 _c$ versus 1/$p^2$ plot in which particles of
a given mass should lie on a straight line. Both the
\v Cerenkov angle and the momentum 
were deduced from each individual ring found on the photon detector.
}
\label{r2p2}
\end{figure}

The RICH can also serve  
 as a tracking device \cite{richstat-roy}. 
This turned out to be very useful in the initial phase of the HERA-B
experiment when the main tracking system was not available. From
the centers of the rings found on the photon detector
by the stand-alone ring search algorithm, 
track directions were deduced.
In the non-bending plane they were matched
with  track candidates from the silicon vertex detector. 
From the deflection in the bending plane the momentum of the particle 
could then be determined.
With both the \v Cerenkov angle and momentum deduced from the data, an 
excellent particle identification was possible even without the main tracking
system, as 
illustrated     in Fig.~\ref{r2p2}.

\section{Summary}

The HERA-B ring imaging \v Cerenkov detector behaves as
expected in all respects. The detector parameters, 33 detected
photons for $\beta =1$ particle, $N_0 = 42$ photons/cm and single
photon resolutions of $\sigma$ = 0.7 mrad (1.0~mrad)
for regions with finer (coarser) granularity,
 are in very good agreement with design values.
Estimates of the basic parameters
obtained from measurements of physical properties of RICH components
agree well with the direct measurements. Therefore, we conclude that
the functioning of the counter is well understood.
Results on particle identification of highly populated events are also
satisfactory and meet the requirements.
Finally, it is worth adding that during the 4 years the detector
has been in operation, not the least degradation of
performance has been 
observed.

\section{Acknowledgments}

We want to express our thanks to the DESY laboratory for the strong 
support since the conception of the HERA-B experiment.
We are especially grateful to the following persons and their groups: 
M.~Bosteels (CERN), J.~Er\v zen (IJS Ljubljana),
K.~Ludwig (DESY), E.~Michel (DESY), 
H.B.~Peters (DESY), P.~Pietsch (DESY), 
M.~Pohl (DESY Zeuthen), J.~Spengler (MPI Heidelberg), and T.~Stoye (DESY).
We acknowledge the support from the U.S. Department of Energy (DOE), 
 Ministry of Education, Science and Sports of Slovenia, 
Funda\c c\~ao para a Ci\^encia e Tecnologia of Portugal,
Spanish CICYT under the  contract AEN99-0483, Internationales B\" uro, J\" ulich, 
 and the  Texas Advanced Research Program.

\end{document}